\documentclass[preprint,12pt,3p]{elsarticle}
\usepackage[english]{babel}
\usepackage{times}
\usepackage{hyphenat}
\usepackage{booktabs}
\usepackage{gensymb}
\usepackage{siunitx}
\usepackage{amsmath}
\usepackage{subfig}
\usepackage[figuresright]{rotating}
\usepackage{multirow}
\usepackage{setspace}
\usepackage{hyperref}
\usepackage[nomarkers]{endfloat}
\usepackage{lipsum}
\usepackage{everysel}
\usepackage{lineno}
\usepackage[dvipsnames]{xcolor}
\usepackage{wrapfig}
\usepackage{adjustbox}
\biboptions{sort&compress}
\usepackage[cmintegrals,cmbraces]{newtxmath}
\usepackage[sfdefault,condensed]{roboto}
\usepackage[T1]{fontenc}

\begin{document}
	
\begin{frontmatter}
	
	\journal{arXiv)}
	
	\title{\Large\textbf{Perspectives on Novel Refractory Amorphous High-Entropy Alloys in Extreme Environments}}
	
	\author[LANL]{M.A. Tunes}
	\cortext[cor]{Corresponding authors.}
	\author[LANL]{H.T. Vo}
	\author[CINT]{J.K.S. Baldwin}
	\author[LANL]{T.A. Saleh}
	\author[LANL]{S.J. Fensin}
	\author[LANL]{O. El-Atwani\corref{cor}}\ead{osman@lanl.gov}

	\address[LANL]{Materials Science and Technology Division, Los Alamos National Laboratory, United States of America}
	\address[CINT]{Center of Integrated Nanotechnologies, Los Alamos National Laboratory, United States of America}
	
	\begin{abstract}
		\onehalfspacing
		
		\noindent Two new refractory amorphous high-entropy alloys (RAHEAs) within the W--Ta--Cr--V and W--Ta--Cr--V--Hf systems were herein synthesized using magnetron-sputtering and tested under high-temperature annealing and displacing irradiation using \textit{in situ} Transmission Electron Microscopy. While the WTaCrV RAHEA was found to be unstable under such tests, additions of Hf in this system composing a new quinary WTaCrVHf RAHEA was found to be a route to achieve stability both under annealing and irradiation. A new effect of nanoprecipitate reassembling observed to take place within the WTaCrVHf RAHEA under irradiation indicates that a duplex microstructure composed of an amorphous matrix with crystalline nanometer-sized precipitates enhances the radiation response of the system. It is demonstrated that tunable chemical complexity arises as a new alloy design strategy to foster the use of novel RAHEAs within extreme environments. New perspectives for the alloy design and application of chemically-complex amorphous metallic alloys in extreme environments are presented with focus on their thermodynamic phase stability when subjected to high-temperature annealing and displacing irradiation. 
		
	\end{abstract}
	
	
\end{frontmatter}
\onehalfspacing

\section{Introduction}
\label{sec:intro}

\noindent An amorphous metallic alloy can be defined by the atomic arrangement: a lack of long-range atomic ordering, or similarly, the absence of long-range translational symmetry \cite{luborsky1983amorphous}. The synthesis of amorphous metallic alloys has been conventionally achieved via (ultra-)rapid solidification of the alloy constituents from either liquid or gas phases. Such rapid cooling prevents the formation of a thermodynamic stable crystalline phase. In a comprehensive review, Luborsky commented that either via deposition or freezing, the transition to solidification occurs so fast that \textit{``the atoms are frozen in their liquid [or vapour] configuration''} \cite{luborsky1983amorphous}. Despite the lack of long-range ordering (LRO) as the major constitutive fact for amorphous metallic alloys, previous research showed evidence that indicated the manifestation of short-range ordering (SRO) as a major characteristic in these alloys  \cite{luborsky1983amorphous,chen1980glassy,jones1973rapid,cargill1975structure,gilman1980metallic}.

Typically formed by the alloying of transition metals (Zr, Hf, Fe, Co etc) with metalloid elements such as B, Si, C, P and Al, transition metal-metalloid alloys with an amorphous structure (herein defined as TMMs) have been reported to exhibit several properties of interest for a wide variety of applications. From a mechanical behavior perspective, TMMs can exhibit high hardness along with surprisingly high tensile strength \cite{luborsky1983amorphous}. Despite the exceptional mechanical behavior, TMMs exhibit soft-magnetic properties that mitigate magnetic loses when compared with crystalline alloys \cite{luborsky1983amorphous,Kronmuller1983}. Additionally, some TMMs are reported to have zero thermal expansion \cite{luborsky1983amorphous,Onn1983} with electrical resistivity that can be four times higher than ferromagnetic iron and its alloys \cite{luborsky1983amorphous,Rao1983}. Resistance to corrosion is another topic where previous research have shown these alloys to possess extraordinary corrosion resistance \cite{luborsky1983amorphous,Davies1983}. 

When in the liquid phase, a metallurgical system can be supercooled with rates on the order of 10$^{6}$ K$\cdot$s$^{-1}$ to form an amorphous solid. In this case, the final material is often a glassy metal alloy, or a bulk metallic glass (BMG) \cite{gilman1980metallic}. Such a discovery within the Au-Si binary system made by Klement, Willens and Duwez in 1960 \cite{klement1960non} caused significant excitement among the worldwide communities of metallurgy and materials science, but amorphous metallic alloys were known before that date by the thin solid film community \cite{gilman1980metallic}. Via physical vapour deposition (PVD) methods such as magnetron-sputtering (MS) or ion-beam sputter deposition (IBSD), high cooling rates can be easily achieved and a wide variety of amorphous materials were reported to be synthesized by supercooling plasmas and gases on a cold substrate. In a different way, solid-solid state reactions were also reported to be a way to synthesize metallic glasses \cite{Schwarz1983,Schwarz1985,Schwarz1986}.

Similar to the complexity of TMMs and BMGs, research efforts within metallurgy have recently focused in the study of complex and concentrated solid solution multicomponent alloys, herein simply referred to as High-Entropy Alloys (HEAs). HEAs bear similarities with both TMMs and BMGs. Both can be considered highly-concentrated metallic alloys, often synthesized via conventional methods with four or more constituents, but HEAs are specially synthesized with (or closer to) equimolar compositions \cite{cantor2004microstructural,cantor2020multicomponent,Bhadeshia2015}. From a thermodynamic point-of-view, HEAs are supposed to be highly stable alloys due to the maximization of configurational entropy, which leads to a lowering of the Gibbs free energy \cite{murty2019high}. Nevertheless, the major constitutive hypotheses for HEAs as a neoteric class of metallic alloy systems within metallurgy are now under intense debate by scientists in many fields \cite{Bhadeshia2015,paul2017comments,schon2018probing,schon2019proof,schon2020complexity,tunes2019investigating,tunes2021comparative,tsai2013sluggish,tsai2017reply,divinski2018mystery,li2013vitro}. 

HEAs can be manufactured using arc-melting or even via powder metallurgy, but many efforts have also focused on using PVD methods. On the latter aspect, a particular field of application of HEAs is in extreme environments such as fission \& fusion nuclear reactors \cite{tunes2018synthesis,tunes2019candidate,elatwani2021helium,Zhang2018}. For these applications, the response of the HEA thin films when subjected to impact of highly-energetic ions or neutrons and their corrosion resistance are of particular interests. The state-of-the-art in the behavior of HEAs under irradiation -- as recently reviewed by Zhang \textit{et al.} \cite{zhang2021tunable} -- highlights that HEAs can be chemically tuned to enhance radiation resistance. All these works relating HEAs with potential application in extreme environments, so far, mainly considered HEAs with crystalline structures. Therefore, the particular development of amorphous HEAs and their concurrent response to energetic particle irradiation is yet pending detailed investigation.

According to the most recent literature \cite{ding2013high,li2013vitro,qi2015soft,muftah2021corrosion}, HEAs in which the final microstructures are amorphous can be categorized either as amorphous high-entropy alloys (AHEAs), bulk metallic high-entropy alloys (BM-HEAs) or even high-entropy bulk metallic glasses (HE-BMGs). The community reports that AHEAs are manufactured in a similar way as TMM alloys or BMGs: rapid quenching from the vapor state (in case of PVD) or liquid phase (in case of conventional metallurgical approaches) \cite{muftah2021corrosion}, but they preserve their particular highly-concentrated multicomponent character.

The purpose of this paper is to discuss the emerging field of AHEAs with particular focus on alloys composed of refractory transition metals: the refractory amorphous high-entropy alloys (herein defined as RAHEAs). Specifically, two RAHEAs within the quaternary WTaCrV and quinary WTaCrVHf metallurgical systems were synthesized using magnetron-sputtering deposition for the first time. After electron-microscopy characterization of both alloys in their as-deposited state, a series of \textit{in situ} Transmission Electron Microscopy (TEM) studies were performed in order to shed light on the alloys' phase stability both under thermal annealing at high-temperatures and under heavy-ion irradiation. We report on the feasibility to manufacture these RAHEAs and also on the fact that phase stability can be attained in specific circumstances, preserving its amorphous nature, thus indicating that this new field of research can fully harness the enterprise of the HEAs for accelerating the discovery of new metallic amorphous materials for applications in extreme environments.

\section{Materials and methods}
\label{sec:matandmet}

\subsection{Synthesis of the alloys using magnetron-sputtering}
\label{sec:matandmet:synthesis}
\noindent Thin solid films of both WTaCrV and WTaCrVHf RAHEAs were prepared using the conventional magnetron-sputtering method from targets with purity levels of 99.99\%. For the quaternary alloy, the magnetron powers were set to 100, 250, 550 and 150 W respectively for W, Ta, Cr and V. In the case of the quinary alloy, the magnetron powers were set to 225, 300, 50, 300 and 50 W respectively for W, Ta, Cr, V and Hf. No intentional heating was applied during deposition of both alloys, therefore, the synthesis was carried out at room temperature. The alloys were deposited onto pure NaCl substrates and their final thicknesses were around 100 nm.

\subsection{Sample preparation and electron-microscopy characterization}
\label{sec:matandmet:characterization}

\noindent After magnetron-sputtering deposition, sample preparation for TEM was performed using the floating method. Substrates with the alloys were immersed in a solution containing ethanol and deionized water (1:2) to deattach the film from the substrate. Then, Mo mesh grids were used to capture the films floating on the solution. With the Mo grids containing both deposited RAHEAs, electron-microscopy characterization was carried out in a FEI Apreo Scanning Electron Microscope (SEM) and in both FEI Tecnai TF30 and FEI Titan 80 Transmission Electron Microscopes (TEM). Energy Dispersive X-ray (EDX) spectroscopy was carried out to estimate the composition of both as-deposited RAHEAs.

\subsection{Thermal annealing}
\label{sec:matandmet:annealing}

\noindent Prior to irradiation, a detailed annealing study was performed in both WTaCrV and WTaCrVHf RAHEAs. The samples were subjected to thermal annealing \textit{in situ} within a FEI Tecnai TF30 TEM operating a field-emission gun at 300 keV. The samples were annealed up to 1173 K in a 10 minutes ramp. After the ramp was attained, the samples were left for 90 min at 1173 K then cooled down to room-temperature in 3 min. For annealing within the TEM, a Gatan 652 double-tilt heating holder was used. Possible modifications in both alloys' microstructures as a result of thermal annealing were monited using the bright-field TEM (BFTEM) and selected-area electron diffraction (SEAD) techniques.

\begin{figure}
	\centering
	\includegraphics[scale=0.75,keepaspectratio]{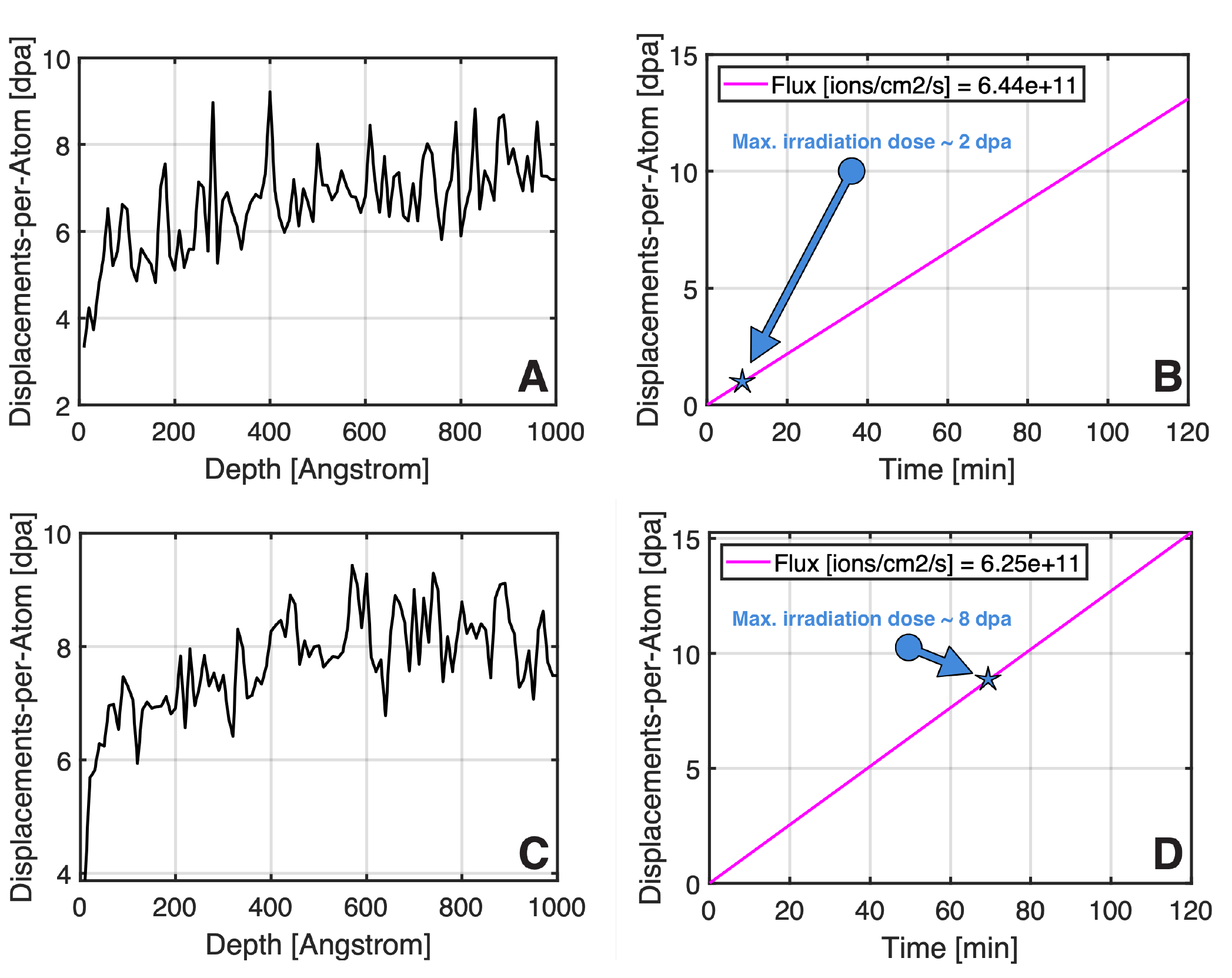}
	\caption{Irradiation conditions for the \textit{in situ} TEM heavy ion experiments using 1 MeV Kr ions. The plots in the figure show the damage profiles and the irradiation dose as a function of the irradiation time for both (A,B) WTaCrV and (C,D) WTaCrVHf RAHEAs. Note: dose in dpa represents an average over the thicknesses (100 nm) of the samples.}
	\label{fig01}
\end{figure}

\subsection{Heavy ion irradiation with \textit{in situ} TEM}
\label{sec:matandmet:irradiation}

\noindent Samples were subjected to dual-beam irradiation \textit{in situ} within a TEM using both 16 keV He$^{+}$ and 1 MeV Kr$^{+}$ ions in the Intermediate Voltage Electron Microscopy (IVEM) facility at Argonne National Laboratory. The quaternary WTaCrV RAHEA was irradiated at room temperature whereas the quinary WTaCrVHf RAHEA was irradiated at 1073 K after 10 min of annealing at 1173 K. For the 1 MeV Kr$^{+}$ case, the alloys were subjected to irradiation fluxes of 6.44$\times$10$^{11}$ (quaternary alloy) and 6.25$\times$10$^{11}$ (quinary alloy) ions$\cdot$cm$^{-2}$s$^{-1}$, respectively. Regarding He irradiations, samples were subjected to an identical flux of around 9.92$\times$10$^{12}$ ions$\cdot$cm$^{-2}$s$^{-1}$. 

Quick-damage calculations using Stopping Range of Ions in Matter (SRIM) and the Kinchin-Pease model were carried out to convert fluence-to-dpa (displacements-per-ion or dpa) and the results are presented in figures \ref{fig01}(A-B) for the quaternary alloy and in figures \ref{fig01}(C-D) for the quinary alloy under only heavy-ion irradiation. Given the fact that all alloy constituents are transition metals, the displacement energy for all the elements was set to 40 eV following a suggestion proposed by Stoller \textit{et.al.} \cite{stoller2013use}. The calculations were performed using the measured compositions of the alloys as-deposited (see table \ref{table_01}) and assuming a thickness of 100 nm. The heavy-ion irradiation set-up causes a similar damage energy profile in both alloys with the only difference being the maximum dose attained: 2 dpa for the quaternary alloy and 8 dpa for the quinary alloy. As for the light-ion irradiation case, the generation of vacancies per ion is between 10-20, therefore, He irradiation was found to cause no significant damage in these samples when compared with the heavy-ion irradiations. Nevertheless, He irradiation was performed to analyze possible formation of He bubbles in the damaged microstructures. The maximum He fluencies for both quaternary and quinary alloys were estimated to be 4.75$\times$10$^{15}$ and 4.16$\times$10$^{16}$ ions$\cdot$cm$^{-2}$, respectively. This corresponds to a dose level of less than 0.2 dpa, and is thus negligible when compared with the radiation damage caused by 1 MeV Kr$^{+}$ ions. Further details on the operational capabilities of the IVEM facility can be found elsewhere \cite{meimei2021situ}.

\begin{table}
	\begin{center}
		\caption{Measured elemental composition of the RAHEAs as-deposited.}
		\label{table_01}
		{\begin{tabular}{lccccc} \toprule
				& \multicolumn{5}{l}{\textbf{Elemental composition [at.\%]}} \\ \cmidrule{2-6}
				\textbf{Alloy} & \textbf{W} & \textbf{Ta} & \textbf{Cr} & \textbf{V} & \textbf{Hf}  \\ \midrule
				WTaCrV & 13.5$\pm$2.5 & 40.9$\pm$3.9 & 26.4$\pm$3.9 & 19.2$\pm$2.5 & --  \\
				WTaCrVHf & 24.0$\pm$0.1 & 39.8$\pm$0.9 & 18.3$\pm$0.6 & 4.6$\pm$0.6 & 13.3$\pm$0.4 \\ \bottomrule
		\end{tabular}}	
	\end{center}
\end{table}

\subsection{Post-irradiation characterization within the analytical electron-microscope}
\label{sec:matandmet:postirradiation}
\noindent After irradiation at the IVEM facility, samples were subjected to post-irradiation screening using a FEI Titan 80-300 TEM which operates a field-emission gun at 300 keV. This instrument is equipped with an Energy Dispersive X-ray (EDX) detector model EDAX Octane Elite T Super for analytical quantification and chemical mapping in the Scanning Transmission Electron Microscopy (STEM) mode.

\section{Results}
\label{sec:results}

\subsection{Characterization of the as-deposited WTaCrV and WTaCrVHf RAHEAs}
\label{sec:results:characterization}

\begin{figure}
	\centering
	\includegraphics[width=\textwidth,height=\textheight,keepaspectratio]{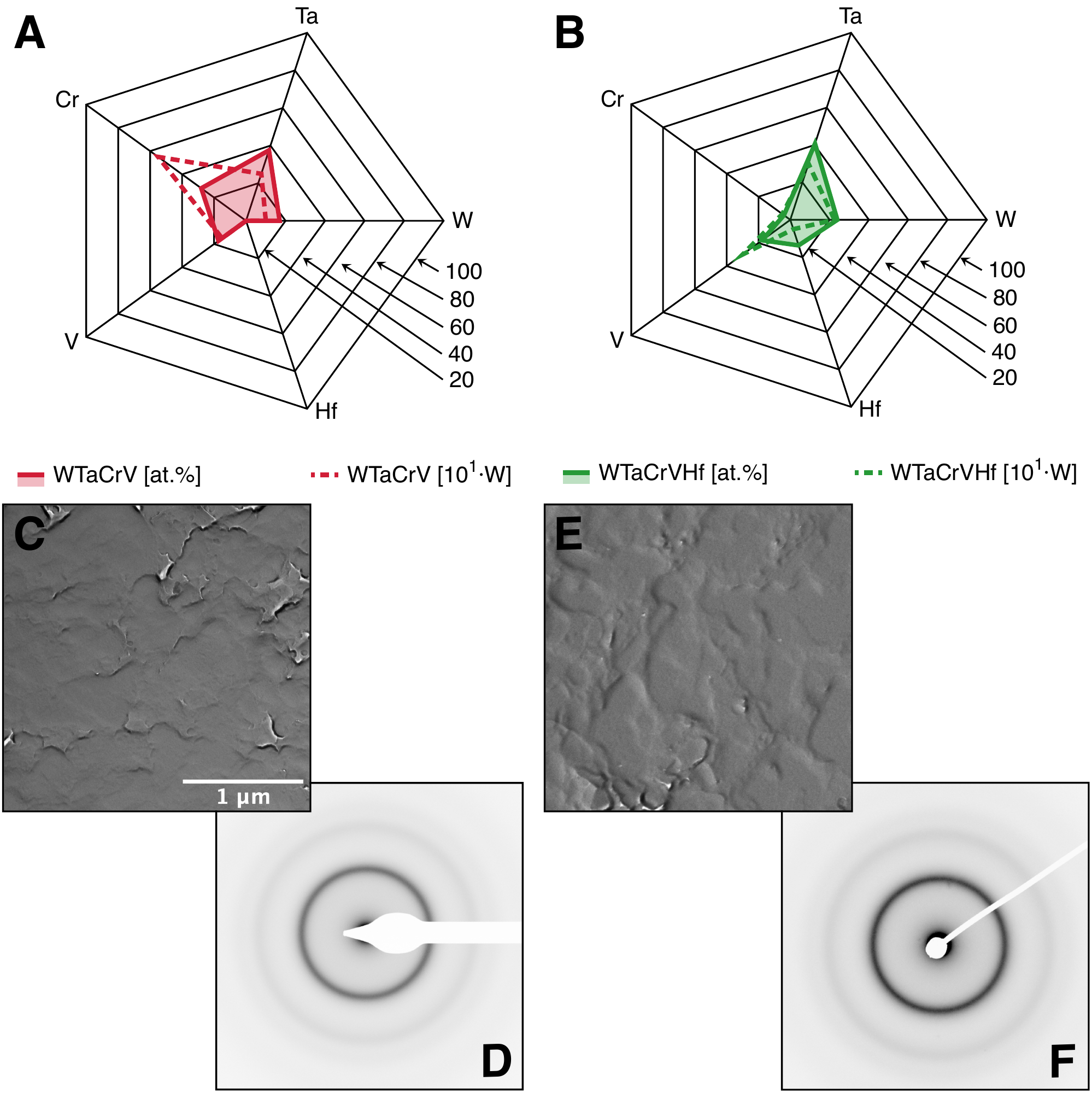}
	\caption{Graphic representation of the relationship between magnetron powers and compositions in a form of spider plots for the (A) WTaCrV and (B) WTaCrVHf RAHEAs as-deposited. The microstructure of both amorphous alloys in the as-deposited condition are shown in the BFTEM micrographs and SAED patterns in (C,D) and (E,F), respectively for the WTaCrV and WTaCrVHf RAHEAs.  Note: the magnetron powers were normalized to 100. The scale bar in C also applied to E.}
	\label{fig02}
\end{figure}

\noindent After magnetron-sputter deposition, the elemental composition of both WTaCrV and WTaCrVHf RAHEAs were measured using SEM-EDX and the results are in figures \ref{fig02}(A) and \ref{fig02}(B), respectively. In these spyder-type plots, the composition of the alloys are compared with the magnetron-sputtering powers (normalized by 100) in order to guide reproducibility in future research. Followed by deposition, the alloys in their as-deposited condition were further investigated within the TEM and the results using the techniques of Bright-Field TEM( BFTEM) and Selected Area Electron Diffraction (SEAD) patterns are shown in the figures \ref{fig02}(C,D) and figures \ref{fig02}(E,F), respectively for the WTaCrV and WTaCrVHf. A characteristic contrast of amorphous materials was observed using BFTEM, \textit{i.e.} the absence of distinct crystalline features (\textit{e.g.} bending contours) was noted. The only major features observed in both alloys under BFTEM contrast was the presence of nanometer-sized holes throughout the samples. However in the quaternary RAHEA, these holes partially disappear upon annealing at high-temperatures, but not under irradiation at room temperature. The collection of diffraction patterns corroborate the BFTEM findings by showing only a collection of amorphous and diffusive ring patterns instead of characteristic crystalline Bragg reflections. Given this preliminary assessment of the as-deposited RAHEAS, a series of investigations on both RAHEAs subjected to both high-temperature annealing and ion irradiation exposure were performed firstly with the WTaCrV (section \ref{sec:results:WTaCrV}) and secondly with the WTaCrVHf (section \ref{sec:results:WTaCrVHf}).

\subsection{WTaCrV RAHEA}
\label{sec:results:WTaCrV}

\noindent The results in this section report on the response of the WTaCrV RAHEA under high-temperature annealing and dual-beam ion irradiation at 293 K of the as-deposited specimens. No irradiation experiment after annealing was performed. A video combining the results herein presented can be found in the supplemental information.

\begin{figure}
	\centering
	\includegraphics[width=\textwidth,height=\textheight,keepaspectratio]{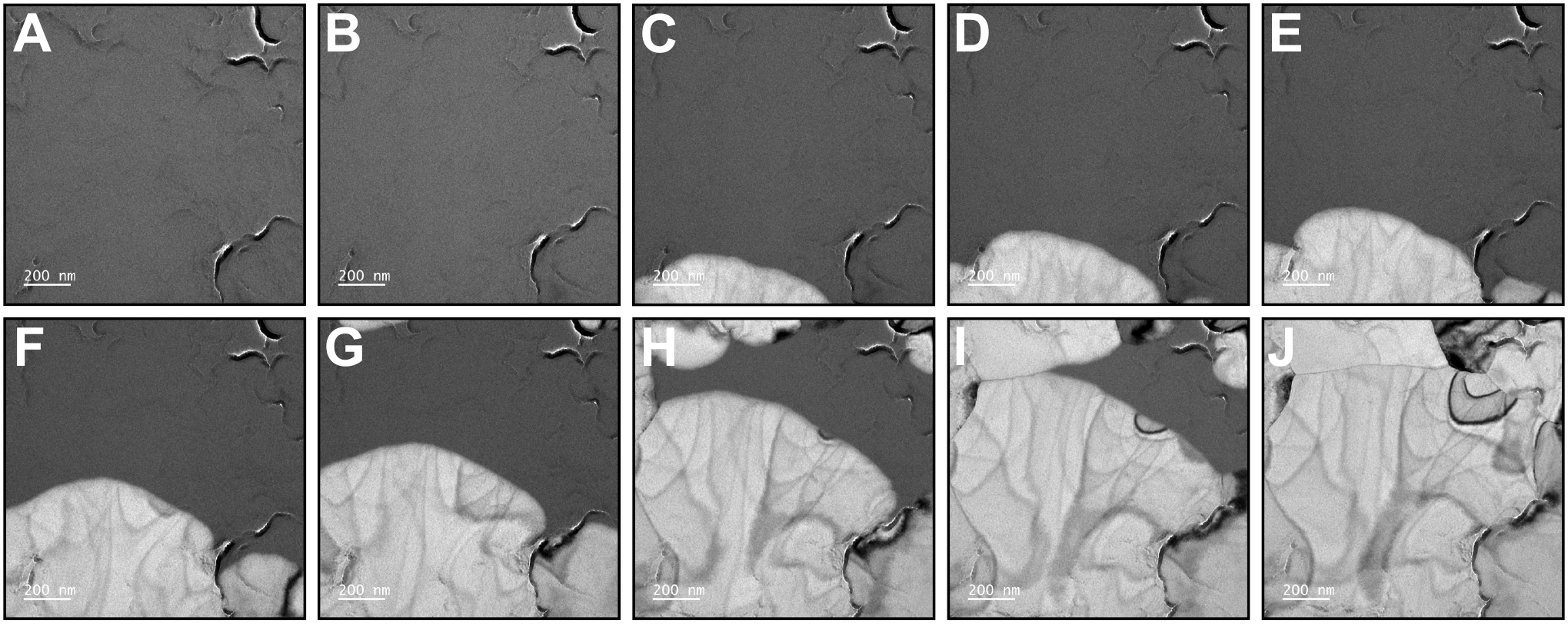}
	\caption{BFTEM micrographs of the \textit{in situ} TEM annealing of the quaternary RAHEA WTaCrV. Micrograph in A represents the amorphous alloy at around 400 K in its pristine amorphous state. Complete crystallization is shown in micrograph J and was observed to take place during 20 s at around 430 K. Micrographs from A to J have 2 s increments. After complete crystallization at 20 s (J), the alloy remained crystalline both up to 1173 K and after quenching to RT. The initial amorphous state was never recovered.}
	\label{fig03}
\end{figure}

The response of the WTaCrV RAHEA to \textit{in situ} TEM annealing is shown in the set of BFTEM micrographs in figure \ref{fig03}(A-J). Initially fully amorphous, the WTaCrV RAHEA experienced ultra-fast crystallization at a temperature of 430 K. The alloy microstructure was observed to be fully crystalline after 20 s of thermal annealing around 430 K as noted in the BFTEM micrograph in figure \ref{fig03}(J). This crystallization is different from the initial amorphous state by the presence of grain boundaries and bending contours, which are characteristic features of crystalline materials when under observation within the TEM. Annealing was performed up to 1173 K, but no significant modifications were observed after full crystallization at 430 K. In addition, after quenching within the TEM vacuum environment, the alloy remained fully crystalline and the initial amorphous condition observed in the as-deposited samples was never restored.

\begin{figure}
\centering
\includegraphics[width=\textwidth,height=\textheight,keepaspectratio]{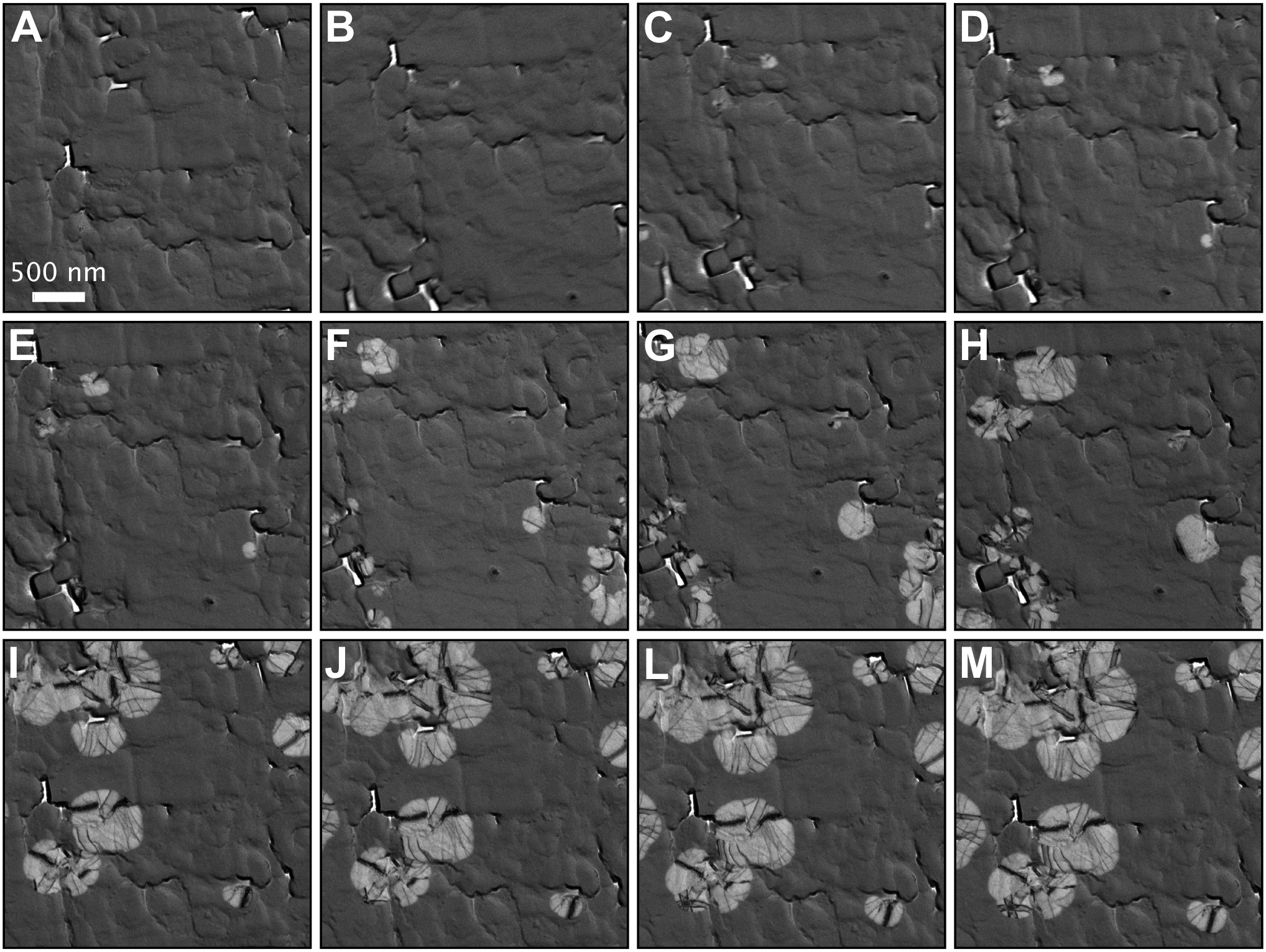}
\caption{BFTEM micrographs recorded during the DB ion irradiation experiments at RT. The quaternary RAHEA WTaCrV in its pristine state is shown in micrograph A taken immediately before irradiation. Irradiation-induced crystallization begins at 0.18 dpa in a form of embryos as shown in micrograph B. Then, several crystalline zones are observed to nucleate and grow upon increasing the irradiation dose as shown in the subsequent micrographs (C-M). Irradiation was intentionally stopped at 2.0 dpa as shown in micrograph M, and the final microstructure was composed of crystalline zones within an amorphous matrix. Notes: the scale bar in micrograph A applies to all micrographs in the figure; micrographs from B to M have 0.18 dpa increments.}
\label{fig04}
\end{figure}

In order to probe the stability of the WTaCrV RAHEA under irradiation, an experiment consisting of 1 MeV Kr$^{+}$ and 16 keV He$^{+}$ dual-beam ion irradiations was performed \textit{in situ} within a TEM. The alloy was irradiated at room temperature in its as-deposited state. The series of BFTEM micrographs in figure \ref{fig04}(A-M) show the results of this experiment. Each micrograph in figure \ref{fig04}(A-M) represents an increment of 0.18 dpa. These micrographs highlight that the while the material was initially amorphous before the irradiation process, controlled irradiation-induced crystallization (IIC) was observed to take place with an increasing irradiation dose. Crystalline zones were observed in the form of embryos -- \textit{i.e.} small nucleation sites -- at 0.18 dpa as shown in figure \ref{fig04}(B), but as the irradiation dose increased, such crystalline zones started to grow, but still in a controlled way. It was also observed that these embryos nucleated and grew radially with respect to its nucleation site center as noted in figure \ref{fig04}(F) taken at 0.9 dpa. At the end of the experiment, figure \ref{fig04}(M) at 2.0 dpa, IIC also promoted coalescence of crystalline zones. At this final dose, the material comprised an amorphous matrix with crystalline zones, therefore, composing a duplex microstructure. It is worth emphasizing that the local BFTEM contrast within the crystalline zones showed distinct bending contours, which corroborates the occurrence of the IIC phenomenon. 

\begin{figure}
\centering
\includegraphics[width=\textwidth,height=\textheight,keepaspectratio]{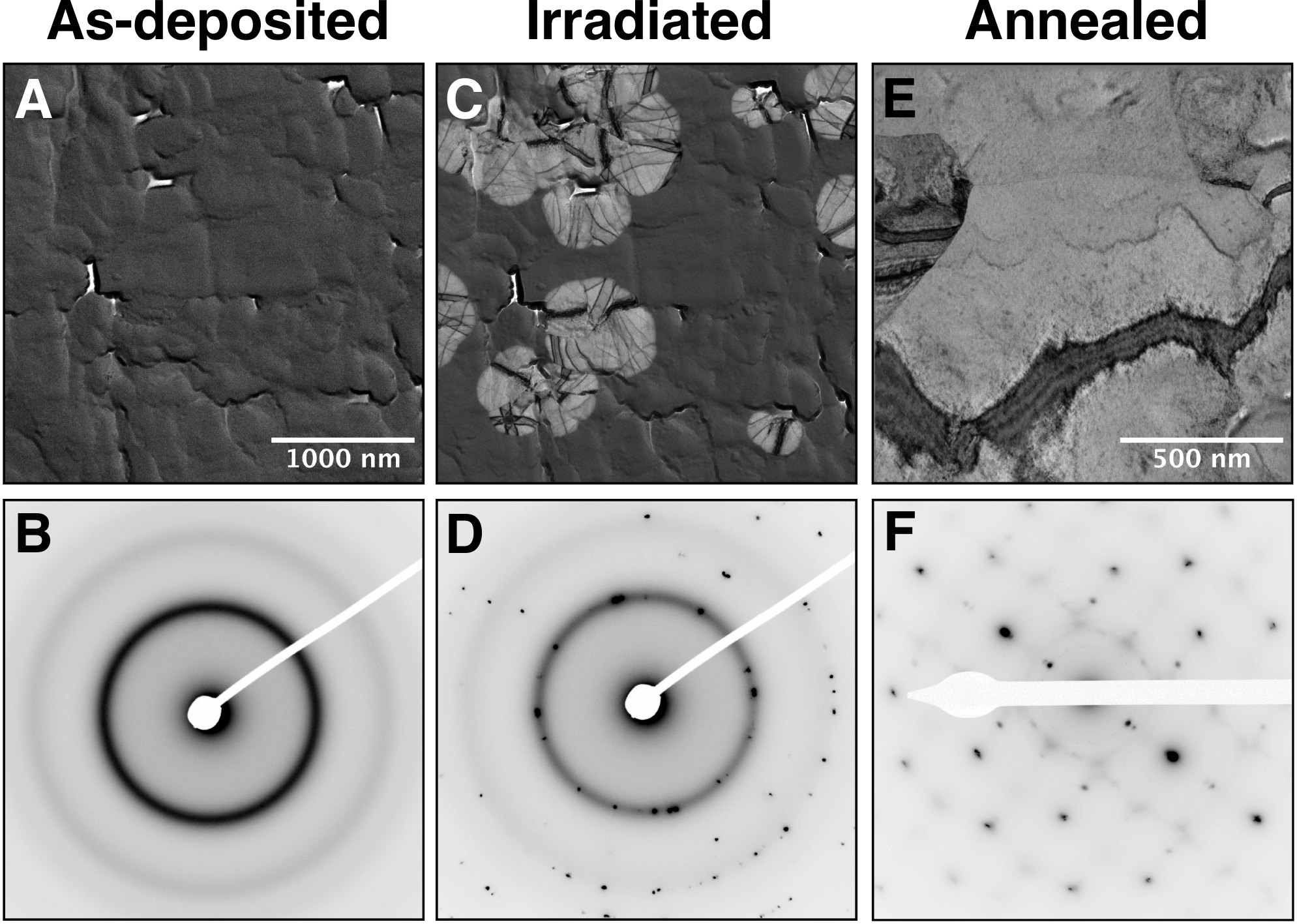}
\caption{A detailed comparison of the quaternary RAHEA WTaCrV using both BFTEM and SAED before irradiation (A,B), after irradiation at RT up to 2 dpa (C,D) and after annealing only up to 1173 K (E,F). Under the irradiated conditions herein investigated, the alloy only partially crystallized after 2.0 dpa: this is particularly noted by analyzing the diffraction pattern in D, where both amorphous rings and crystalline Bragg reflections are noted. Conversely, thermal annealing resulted in full crystallization at a temperature around of 430 K. Irradiation was discovered to be a way to perform controlled crystallization in this alloy. Note: the scale bar in A also applies to C.}
\label{fig05}
\end{figure}

Both irradiated and annealed samples were post-characterized within a TEM to shed light on the observed differences in the microstructure. Figure \ref{fig05}(A-F) shows a collection of BFTEM micrographs and SAED patterns obtained from the WTaCrV RAHEA in its as-deposited, irradiated up to 2.0 dpa and annealed up to 1173 K. The main differences between dual-beam ion irradiation and thermal annealing can now be underlined. It is clear that crystallization happens in an uncontrolled way as a result of solely thermal annealing. Full crystallization was observed at a temperature of only 430 K, which is low when the melting points of the alloy's constituents are taken in consideration. Conversely, irradiation at RT indicated that the phenomenon of IIC takes place, but the amount of crystallization could be well controlled as a function of the irradiation dose. This is particularly interesting considering that the activation of nucleation and growth mechanisms normally results in crystallization in an uncontrollable manner \cite{hollomon1953nucleation}. Amorphization of the ion-beam induced crystallized areas was not observed, and as such upon cooling during the thermal annealing experiments. This suggests that an irreversible amorphous-to-crystalline phase transformation takes place.

\subsection{WTaCrVHf RAHEA}
\label{sec:results:WTaCrVHf}

\noindent The results in this section report on the response of the WTaCrVHf RAHEA under high-temperature annealing and dual-beam ion irradiation at 1073 K. A video combining the results herein presented can be found in the supplemental information.

\begin{figure}
	\centering
	\includegraphics[width=\textwidth,height=\textheight,keepaspectratio]{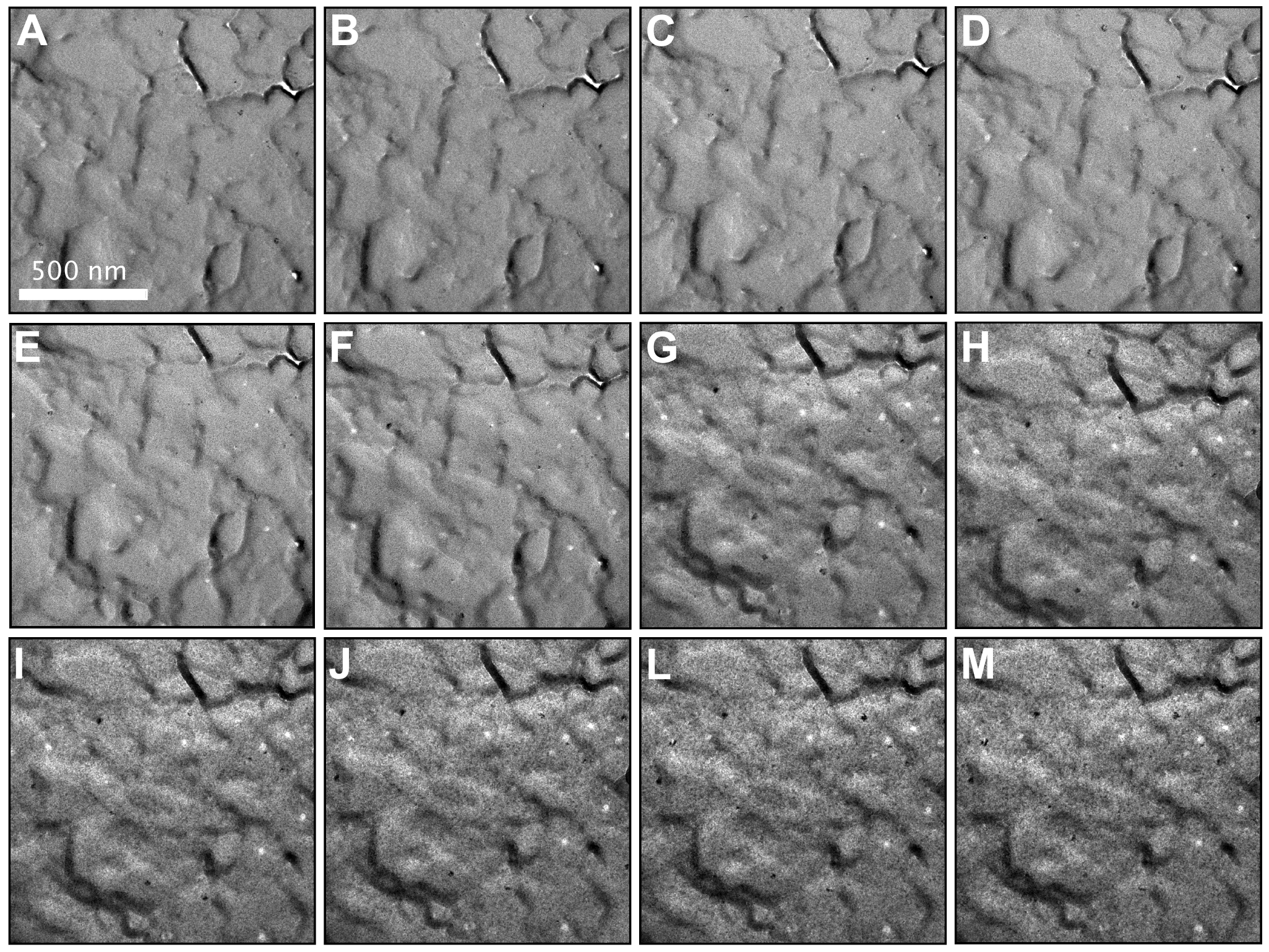}
	\caption{BFTEM micrographs recorded during the dual-beam ion irradiation experiments at 1073 K. The quinary WTaCrVHf RAHEA after 10 min of annealing at 1173 K and before irradiation is shown in micrograph A. At the end of the experiment at 8 dpa, as shown in micrograph M, ion-induced cratering and irradiation-induced precipitation were observed. Notes: the scale bar in micrograph A applies to all micrographs in the figure; micrographs from B to M have 0.73 dpa increments.}
	\label{fig06}
\end{figure}

Identical experiments were also performed in the WTaCrVHf RAHEA. However, no crystallization was observed in this quinary amorphous alloy upon annealing. This particular result is exhibited in the BFTEM micrograph in figure \ref{fig06}(A) which was recorded after 10 min of annealing at a temperature of 1173 K. Thus, the WTaCrVHf RAHEA exhibited thermal stability under high-temperature annealing, which is in contrast to the results for the quaternary WTaCrV RAHEA. It is worth emphasizing that both alloys were deposited under similar magnetron-sputtering conditions with only the composition and the magnetron-sputtering powers as the major differences between them. Thus, thermal stability up to 1173 K under high-temperature annealing was unexpected in the WTaCrVHf RAHEA.

Given the thermal stability in the WTaCrVHf RAHEA, the dual-beam ion irradiation experiments were performed at higher doses and at a higher temperature (1073 K) as opposed to the experiments performed in the quaternary WTaCrV RAHEA. The objective was to probe whether the alloy shows phase stability under irradiation. The microstructural response of the quinary RAHEA under dual-beam irradiations at high-temperature is shown in the set of BFTEM micrographs in figure \ref{fig06}(B-M). Each micrograph has an increment of 0.73 dpa, resulting in a total dose of 8 dpa at the end of the experiment. BFTEM contrast in these micrographs indicates that the amorphous structure was preserved, suggesting matrix stability throughout the experiment, but nanometer-sized round-shaped precipitates became noticeable at a dose around of 4.4 dpa as shown in figure \ref{fig06}(G). BFTEM contrast arising from these precipitates were clearer at the dose of 8.0 dpa as exhibited in figure \ref{fig06}(M). Ion-induced cratering was also observed to nucleate and grow upon increasing the dose. These became noticeable at 2.9 dpa as shown in figure \ref{fig06}(E). These craters are of strong white contrast monitored under BFTEM, suggesting mass-thickness differences between the crater and matrix. 

\begin{figure}
\centering
\includegraphics[width=\textwidth,height=\textheight,keepaspectratio]{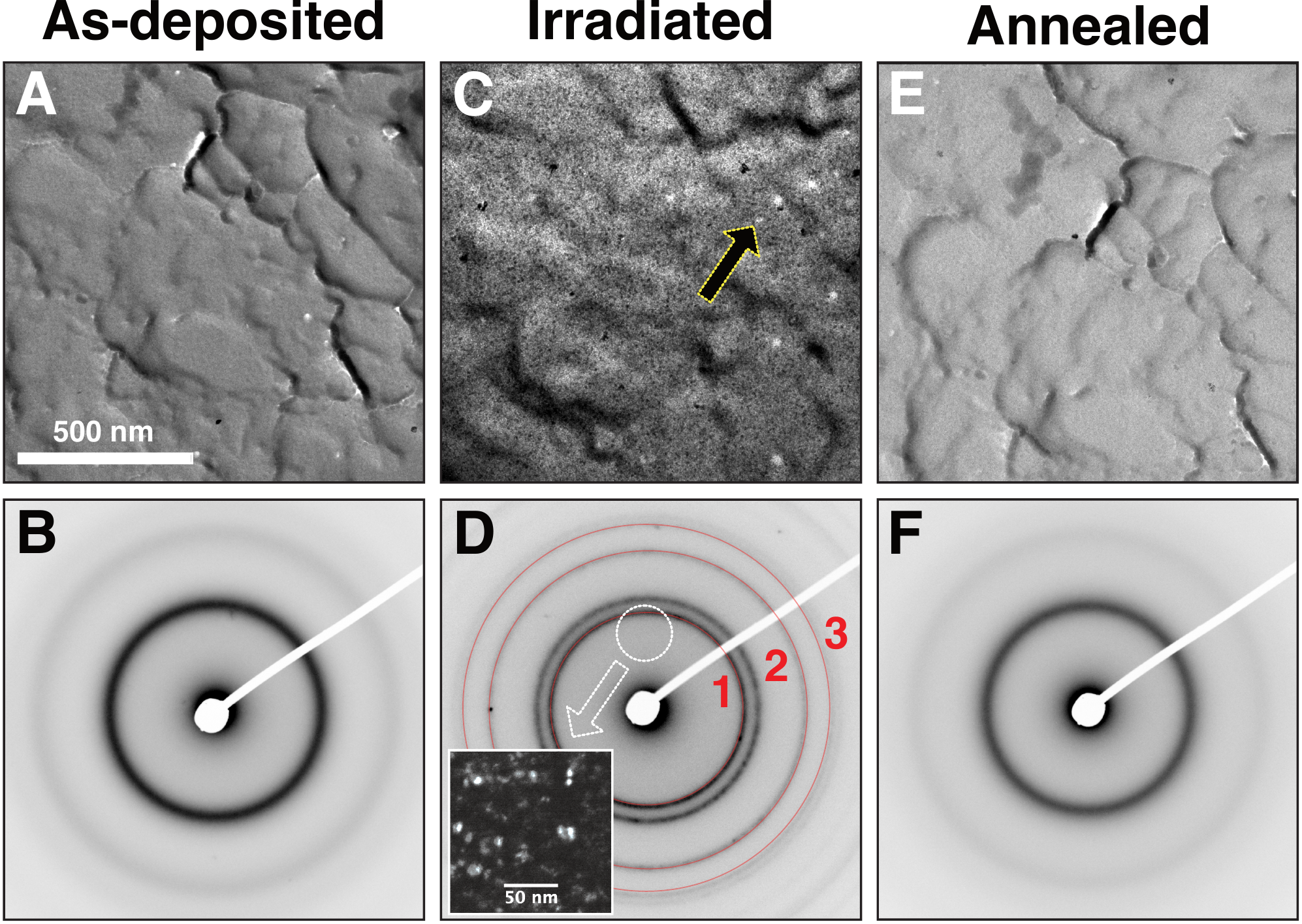}
\caption{A detailed comparison of the quinary RAHEA (A,B) before irradiation, (C,D) after irradiation at 1073 K up to 8 dpa and (E,F) after 10 min of annealing at 1173 K. Under the studied conditions, the alloy suffered irradiation-induced precipitation of fine nanometer-sized particles with crystalline nature as shown in the DFTEM inset in D. The new Debye-Scherrer rings that appeareed after irradiation in the SAED pattern in D were indexed to be pure HCP Hf (see table \ref{table_02}). In addition to precipitation, ion-induced cratering as pointed by the arrow in C. Note: the scale bar in micrograph A also applies to B and C.}
\label{fig07}
\end{figure}

Post-irradiation and -annealing characterization was performed in the quinary WTaCrVHf RAHEA within the TEM. A set of BFTEM micrographs and SAED patterns are presented in figure \ref{fig07}(A-F). Initially amorphous as noted in micrographs \ref{fig07}(A,B), fine nanometer-sized rounded-shaped precipitates were observed to form as a result of the dual-beam irradiation as shown in micrographs \ref{fig07}(C,D). The Dark-Field TEM (DFTEM) inset in figure \ref{fig07}(D) reveals the crystalline nature of such precipitates. Using data available \cite{romans1965transformation}, indexing of the extra-rings observed in the SAED pattern after irradiation in figure \ref{fig07}(D) indicates these precipitates are pure Hf HCP as shown in table \ref{table_02}. Ion-induced cratering -- shown by the arrow in figure \ref{fig07}(C) -- was another type of extended defected formed in the microstructure of the WTaCrVHf RAHEA. Nevertheless, thermal annealing caused no modifications within the alloy's microstructure as shown in micrographs \ref{fig07}(E,F). In summary, the WTaCrHfV RAHEA exhibited resistance to both annealing and irradiation, however, in the latter case, radiation-induced precipitation was observed within the amorphous matrix.

\begin{table}
	\begin{center}
		\caption{Measured and reference d-spacings for the quinary RAHEA WTaCrVHf after irradiation.}
		\label{table_02}
		{\begin{tabular}{ccccc} \toprule
				\textbf{Ring {[}\#{]}} & \textbf{Measured d-spacing {[\AA]}} & \textbf{Ref. d-spacing of pure HCP Hf {[\AA]}} \cite{romans1965transformation} & \textbf{Miller-Planes} & \textbf{I/Imax} \\
				1                      & 2.52                                 & 2.53                                  & (002)                  & 100\%           \\
				2                      & 1.56                                 & 1.59                                  & (110)                  & 46.3\%         \\
				3                      & 1.34                                 & 1.35                                  & (112)                  & 33.9\%  \\
				\bottomrule      
		\end{tabular}}
	\end{center}
\end{table}

\subsection{Local chemistry and thermodynamic phase stability under extremes}
\label{sec:results:localchemistry}

\begin{figure}
	\centering
	\includegraphics[width=\textwidth,height=\textheight,keepaspectratio]{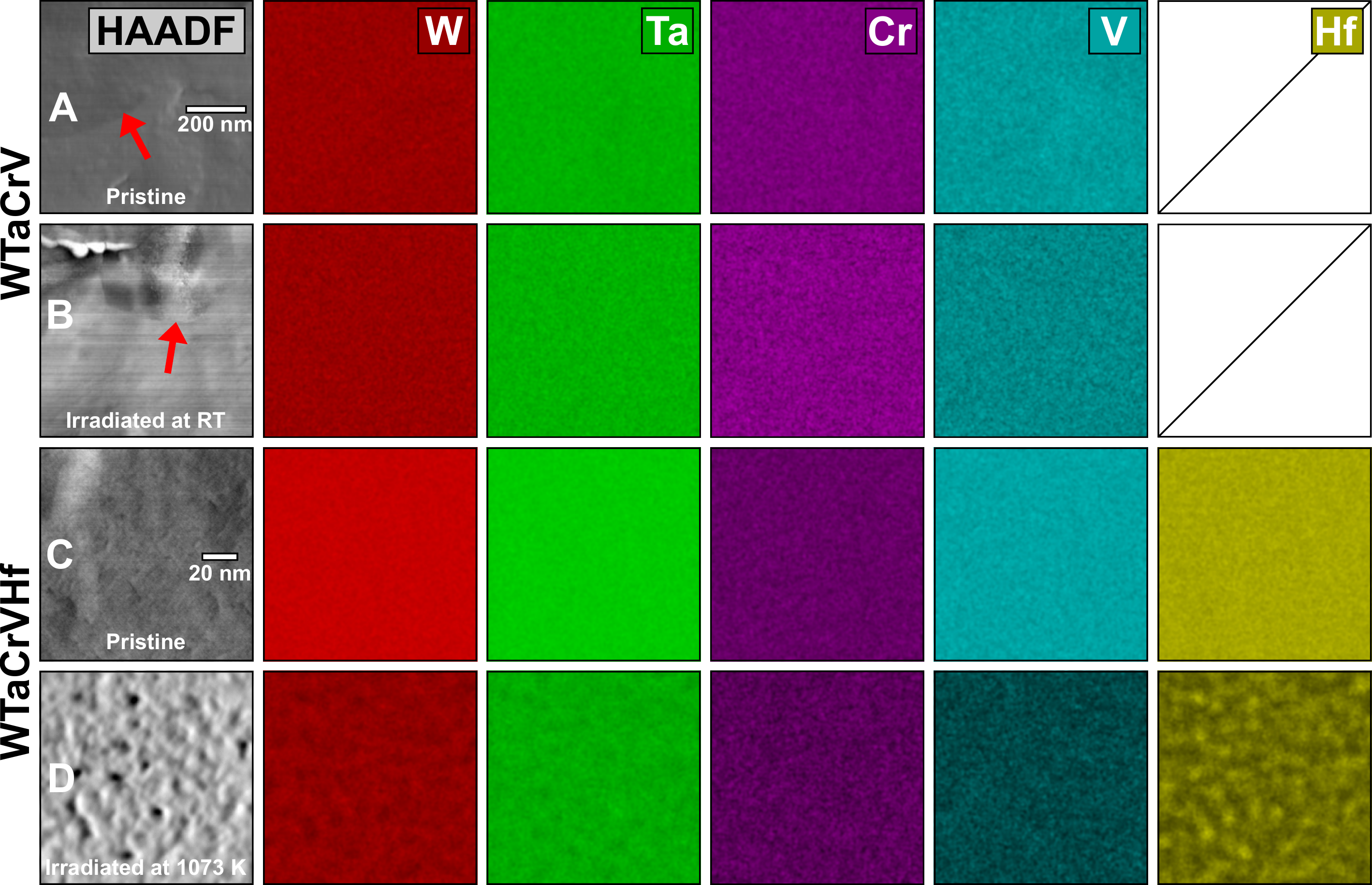}
	\caption{Local chemical analysis of both WTaCrV and WTaCrVHf RHEAs before (A,C) and after (B,D) irradiation. The micrograph in A shows the pristine WTaCrV RHEAs after two months of storage at RT in a desiccator: crystalline areas were observed to be similar to those formed under irradiation (red arrows in A and B). For the quaternary alloy, chemical analysis of both crystalline and amorphous areas before and after irradiation (A,B) revealed neither chemical instabilities nor segregation, whereas nanometer-sized precipitates rich in Hf were observed (D) to form after irradiation at 1073 K on the WTaCrVHf RHEA up to 8 dpa, but the amorphous matrix was preserved (see in Fig. \ref{fig07}).}
	\label{fig08}
\end{figure}

\noindent In order to evaluate possible local chemistry changes as a result of the dual-beam irradiations both WTaVCr and WTaVCrHf RAHEAs were subjected to post-irradiation analysis using STEM-EDX. Figure \ref{fig08}(A-D) shows a series of EDX maps obtained from the as-deposited alloys (referred as pristine) and irradiated at RT for the quaternary alloy and at 1073 K for the quinary alloy. 

For the WTaVCr, the as-deposited sample investigated using STEM-EDX was previously stored for two-months in its pristine form in a desiccator. The High-Angle Annular Dark-Field (HAADF) micrographs in figure \ref{fig08}(A) shows that crystalline zones were observed in this sample without any heat-treatment or irradiation. However, no local chemistry alterations were detected between the amorphous and crystalline zones. Conversely, fine nanometer-sized Hf-rich precipitates were observed to form in the WTaVCrHf RAEHA as a result of irradiation as shown in the EDX maps in figure \ref{fig08}(D). These observations suggest that for the quaternary alloy, the amorphous-to-crystalline phase transformation takes place without any detectable solid-state diffusion, whilst an amorphous matrix is preserved for the quinary alloy, but Hf is seemly diffuses out from the amorphous matrix and forming crystalline nanoprecipitates.

\section{Discussion}
\label{sec:discussion}

\noindent The formation of WTaCrV and WTaCrVHf amorphous alloys under magnetron-sputtering conditions can be understood in the following way. Metallurgically, amorphous metallic alloys are generally synthesized via quenching at fast cooling rates from the liquid state. These rates are sometimes reported to be in the order of 10$^{6}$ K$\cdot$s$^{-1}$ \cite{gilman1980metallic}. Such a fast quench prevents crystallization by hindering atomic mobility such that the atoms within the liquid state are  ``frozen-in'' in the solid state \cite{gilman1980metallic,luborsky1983amorphous}. Obviously no liquid state is present within the magnetron-sputtering conditions for thin film deposition, but a vapour or ionized plasma phase instead is present. When bombarded target atoms become ionized within the magnetron-sputtering chamber, they tend to deposit on a substrate surface where Gibbs thermodynamic potential will dictate the mechanisms of nucleation and growth of embryos \cite{hollomon1953nucleation} on an interface, \textit{i.e.} heterogenous nucleation \cite{turnbull1950kinetics}. Upon coalescence and growth of embryos, crystallization takes place onto the substrate. Thus, the deposition phase transition is very similar to phase transition during solidification from a liquid state. It is worth emphasizing that quenching rates under magnetron-sputtering conditions are reported to be quasi-infinite, sometimes on the order of 10$^{11}$ K$\cdot$s$^{-1}$ for some amorphous metallic alloys \cite{williams1984corrosion}. 

Several different metallic alloy systems were reported to be amorphous under magnetron sputtering \cite{cargill1975structure}, but HEAs can be considered a very particular case. The thermodynamic barrier that the metallic system must overcome for nucleation of crystalline embryos is described by the following equation \cite{hollomon1953nucleation}:

\begin{equation}
	r^{*} \propto -\frac{2\gamma}{\Delta G_{v}}
\end{equation}

Where $r^{*}$ is the critical radius for embryo nucleation (spherical) and subsequent growth, $\gamma$ is the surface energy of the alloy's system and ${\Delta G_{v}}$ as the volumetric Gibbs free energy. During deposition, embryos with $r<<r^*$ will tend to vanish whereas those with $r>>r*$ will nucleate, survive and grow. In HEAs, the Gibbs free energy of the system tends to be largely negative, as these alloys are designed with highly-concentrated compositions, which results in the maximization the configuration entropy of the system. In fact, recent experimental evidence on deposition of thin films within the Fe--Cr--Mn-Ni system has validated the equation (1) for non-refractory HEAs systems \cite{tunes2018synthesis}: the deposition under identical conditions of both an equimolar and non-equimolar thin solid films lead to a smaller grain size than the latter film, pointing to the fact that smaller Gibbs free energies -- characteristic of high-entropy materials \cite{cantor2020multicomponent} -- will reduce the critical radius barrier for crystallization of HEA films, leading to nanocrystalline microstructures.

The mechanism of crystallization of refractory HEAs thin films can be rather complex as compared to non-refractory systems. This is the case for both W--Ta--Cr--V and W--Ta--Cr--V--Hf systems herein under investigation. Movement of adsorbed atoms within a deposition substrate is a required condition for the formation of a thin solid film. A model developed by Movchan and Demschichin \cite{movchan1969rost}, and later refined by Thornton \cite{Thornton1986} (the Movchan-Demschichin-Thronton zone model, or simply MDT zone model), indicate on a dependence between the homologous temperature deposition -- \textit{i.e.} T/T$_m$ where T is the deposition temperature of the substrate and T$_m$ the melting point of the alloy -- with the final microstructure of the deposited film. Herein, both WTaCrV and WTaCrVHf were deposited at room-temperature (298 K). Although the T$_{m}$ of both alloys were not measured due to experimental limitations, a simple average calculation indicates a T$_m$ of approximately 2837 and 2770 K for the quaternary and the quinary alloys, respectively. This yields a homologous temperature for both films on the order of T/T$_m$ $\approx$ 0.10, which can be considered low within the MDT zone models. Films deposited with T/T$_m <$ 0.3 are known to be in zone 1 where the microstructure is often reported to be either composed of large columnar nanocrystalline grains with voided boundaries or completely amorphous \cite{thornton1986microstructure,movchan1969rost}. Therefore, the formation of the present RHEAs herein studied may occur considering the inherent lower atomic mobilities of their refractory metals constituents at low temperatures \cite{neumann1990self} (resulting in hindered solid-state diffusion at the substrate interface) in combination with low deposition temperatures and the specific set of elemental compositions here reported (leading to complex barrier for embryo nucleation).

It is very interesting that these RAHEAs, although very similar, exhibit different responses under irradiation and high-temperature testing. In the case of the quaternary WTaCrV RAHEA, \textit{in situ} TEM annealing up to 1173 K showed full recrystallization was observed at 430 K (figure \ref{fig03}) -- which can be considered very low temperature for the system \textit{i.e.} T/T$_{m} \approx$ 0.17 -- whereas \textit{in situ} TEM ion irradiation at room-temperature indicated that the crystallization phenomenon can be controlled as a function of the irradiation dose (figure \ref{fig04}). The evidence indicates that the WTaCrV fully crystallized at low temperature suggests that the alloy in its as-deposited condition quickly solidified from an unstable thermodynamic condition. Upon heating at very low temperatures, the alloy is driven to a more stable thermodynamic configuration, which is a body-centered cubic (BCC) crystalline structure. This is also confirmed by the fact that the amorphous configuration was never restored upon quenching, \textit{i.e.} pointing to an irreversible phase transformation. 

Still in the WTaCrV RAHEA, it is worth noting that there is no compositional difference between the amorphous and crystalline phases, thus indicating that for this alloy system, the amorphous-to-crystalline phase transformation either triggered by room-temperature irradiation or low-temperature annealing is somehow very complex. Solid-state diffusion can take place during such a phase transformation and this mechanism is described in the literature as a reconstructive diffusional transformation \cite{soffa2014physical,buerger1948role,buerger1951crystallographic}. In this transformation type, atoms from the parent matrix phase are disassembled to form a new phase via atom-by-atom assembling. The crystallization of the amorphous WTaCrV alloy driven either by annealing or irradiation (both at low-temperatures) suggests that solid state diffusion can take place at low-temperatures, even considering that intrinsic mobilities of the alloy constituents can be low at such temperatures \cite{tunes2019microstructural}. These facts are due to significant phase modifications that were observed to take place under the testing conditions reported in this work.  In addition, the alloy in as-deposited state was left for two months in a desiccator and crystalline zones -- similar to those observed under irradiation -- were observed without heating or irradiation. This reinforces the argument that the WTaCrV RAHEA after deposition is in an unstable thermodynamic state.

Different results were obtained with the quinary WTaCrVHf RAHEA tested under identical conditions. The quinary alloy was found the be thermodynamically stable under annealing at 1173 K as the initial amorphous phase did not undergo phase transformation solely as a result of thermal annealing (figure \ref{fig06}A). Irradiation also indicated higher stability of the amorphous phase tested up to a dose of 8 dpa (figure \ref{fig06}M) at 1073 K when compared with the quaternary alloy. Nevertheless, a major irradiation effect was detected to initiate at a dose around 4.38 dpa: the precipitation of nanometer-sized Hf-rich precipitates, confirmed both with SAED pattern indexing and DFTEM as shown in figure \ref{fig07}(C,D), and STEM-EDX as shown in figure \ref{fig08}D. Although the radiation-induced precipitation of crystalline Hf was observed to take place, the matrix remained amorphous up to the end of the irradiation experiment and after the formation and growth of Hf precipitates, no further radiation effect was observed in this alloy upon increasing the dose. In contrast to the WTaCrV RAHEA, it has been herein demonstrated that the addition of Hf to the W--Ta--Cr--V quaternary system tailored the irradiation response of the alloy, thus resulting in an increase of its thermodynamic stability under both high-temperature annealing and high-dose irradiation.

Irradiations in the WTaCrVHf RAHEA indicated that the as-deposited alloy was found to be in a different thermodynamic state in comparison with the WTaCrV RAHEA. IThe precipitation of Hf that initiates at a dose of 4.38 dpa indicates a deleterious effect to the overall alloy's response to irradiation. However, the reassembling of nanoprecipitates resisting radiation has been recently idealized by some authors within the radiation effects community as a new mechanism to enhance the radiation response of functional materials (ceramics and metallic alloys) to be applied in extreme environments \cite{tunes2020prototypic,tunes2021deviating,zhang2022reassembled}. This happens due to the fact that a high areal/volumetric density of precipitates may act as sinks for irradiation induced defects in the form of free volume in the case of an amorphous alloy, thus removing from amorphous solid solution excess free volume that could evolve to form severe damage (e.g. large cavities and bubbles). In this sense, the radiation-induced nanoprecipitation of Hf in the WTaCrVHf RAHEA can be viewed as a self-healing mechanism, resulting in an enhanced radiation tolerance for this quinary amorphous alloy.

The design of materials for extreme environments starts with the discovery of new metallic alloys or high-performance ceramic materials. For historical reasons, these are of crystalline nature. Whilst in-service in extreme environments, the crystalline structures are constantly subjected to the formation of defects at the atomic level. That is the case, for example, in potential materials that will fit future nuclear fusion reactors \cite{knaster2016materials}, where highly energetic fast neutrons and He nuclei impinge into their crystalline structure causing cascade of point defects that may evolve and form extended defect structures such as dislocation loops, cavities and bubbles, precipitation and a wide variety of phase transformations. These extended defects causes accelerated degradation in potential fusion materials, precluding both the prototypic development and the commercialization of fusion power. Nevertheless, the results obtained in this paper for the WTaCrVHf RAHEA indicates a potential revolutionary perspective for the field of radiation effects in materials: in an amorphous metallic alloy, the fundamental notion of crystalline defects does not exist. Displacement damage defects -- such as dislocation loops -- do not exist in an RAHEA. Therefore, radiation damage effects in an amorphous metallic alloy can result in the formation of cavities via impurity damage within the free volume and/or thermodynamic phase instabilities, such as crystallization as observed here for the quaternary WTaCrV RAHEA. However, the quinary WTaCrVHf RAHEA has not exhibited neither of displacement damage defect nor cavities or bubbles, indicating on a high radiation tolerance within the boundary conditions investigated in this work.

The perspectives for use of amorphous alloys in extreme environments are highly counter-intuitive. In fact most reports within the radiation effects and nuclear materials communities were actually focused on studying the amorphization process of crystalline materials as a result of irradiation. This is historical for example in the study of ceramic and semiconductor materials \cite{edmondson2009amorphization,sickafus2000radiation}. As demonstrated in this paper, the spectrum of amorphous RHEAs offer a suitable scenario where the chemical complexity can be tunable for achieving high amorphous phase stability in extreme conditions. The possibility to induce reassembly of precipitates also is a new strategy to possibly enhance the radiation response of an amorphous alloy when under irradiation. It is worth emphasizing again that the notion of crystalline defects induced by irradiation (e.g. dislocation loops, stacking faults) is absent in an amorphous system with only the free volume being responsible to respond to the generation of radiation effects. Therefore, the design of an amorphous alloy with phase stability under irradiation and high-temperature annealing arise from the results obtained in this research as new major alloy design criteria for this exciting field of research for materials under extreme conditions.

\section{Conclusion}
\label{sec:conclusion}

\noindent Two new refractory amorphous high-entropy alloys within the quaternary W--Ta--Cr--V and quinary W--Ta--Cr--V--Hf systems where herein synthesized and tested under high-temperature annealing and under dual-beam irradiations using \textit{in situ} TEM experimentation. It was demonstrated that by tuning the chemical complexity of the alloys with addition of Hf in the quaternary system, phase stability both under annealing and irradiation was achieved. This research demonstrates the feasibility to design new complex amorphous metallic alloys with enhanced radiation resistance for use in extreme environments. The thermodynamic phase stability of the quinary WTaCrVHf RAHEA composes the major criteria when the use of amorphous alloys is proposed for both high-temperature and irradiation environments. Reassembling of nanoprecipitates as a result of irradiation was also herein firstly detected to be related with such enhanced radiation resistance. Further research is required to assess different refractory and non-refractory amorphous HEAs systems under irradiation to understand the role of chemical complexity in the stability of amorphous structures. In addition for the future, the mechanical properties of the most-promising amorphous alloys should be tested and compared with conventional materials currently in-service in extreme environments.

\section*{Acknowledgments}

\noindent Research presented in this article was supported by the Laboratory Directed Research and Development (LDRD) program of Los Alamos National Laboratory primarily under project number 20200511ECR (led by OEA). MAT acknowledges support from the LDRD program 20200689PRD2. This work was supported by the U.S. Department of Energy, Office of Nuclear Energy under DOE Idaho Operations Office Contract DE-AC07- 051D14517 as part of a Nuclear Science User Facilities experiment.

\section*{Disclosure statement}

\noindent The authors declare no conflict of interest.

\section*{Data availability}

\noindent All the data is already presented in the manuscript and supplemental materials. Raw data is available upon request.

\section*{References}
\bibliographystyle{elsarticle-num}
\bibliography{biblibrary.bib}

\newpage
\tableofcontents

\end{document}